\documentclass[%
superscriptaddress,
twocolumn,
amsmath,amssymb,
aps,
]{revtex4-2}

\usepackage{graphicx}
\usepackage{dcolumn}
\usepackage{bm}
\usepackage{hyperref}
\usepackage{physics}
\usepackage{mathtools}
\usepackage{mathrsfs}
\usepackage{siunitx}

\newcommand{\ha}{\hat{a}^{\hphantom{\dagger}}} 
\newcommand{\had}{\hat{a}^\dagger} 
\newcommand{\asic}{a\mathrm{SiC}}
\newcommand{\asio}{a\mathrm{SiO_2}}
\newcommand{\asi}{a\mathrm{Si}}

\usepackage{xcolor}
\definecolor{amber}{rgb}{1,0.49,0}
\definecolor{darkgreen}{rgb}{0,0.55,0}

\DeclareFontFamily{U}{mathx}{\hyphenchar\font45}
\DeclareFontShape{U}{mathx}{m}{n}{<-> mathx10}{}
\DeclareSymbolFont{mathx}{U}{mathx}{m}{n}
\DeclareMathAccent{\widebar}{0}{mathx}{"73}

\usepackage[normalem]{ulem}
\definecolor{tangerine}{rgb}{0.944,0.522,0}
\definecolor{verde}{rgb}{0.,0.6,0}
\definecolor{rosso}{rgb}{0.9,0.0,0.2}
\definecolor{magenta}{rgb}{0.9,0.2,0.9}

\newif\ifhighlight

\newcommand{\highlight}{\highlighttrue}
\highlight
\newcommand{\editor}[2]{%
  \expandafter\newcommand\csname #1note\endcsname[1]{%
    \textcolor{#2}{(\textbf{#1:} ##1)}}%
  \expandafter\newcommand\csname #1\endcsname[1]{%
    \ifhighlight\textcolor{#2}{##1} \else ##1\fi}%
  \expandafter\newcommand\csname #1cancel\endcsname[1]{%
    \ifhighlight\textcolor{#2}{\sout{##1}}\fi}%
  \expandafter\newcommand\csname #1change\endcsname[2]{%
    \ifhighlight\textcolor{#2}{\sout{##1} ##2}\else ##2\fi}%
  \newenvironment{#1text}{\ifhighlight\color{#2}\fi}{\color{black}}
}

\editor{SB}{tangerine}
\editor{PP}{verde}
\editor{resub}{rosso}
\editor{AF}{blue}
\editor{CAVEAT}{red}

\usepackage{orcidlink}
\begin{document}

\title{Hydrodynamic finite-size scaling
of the thermal conductivity in glasses}

\author{Alfredo Fiorentino,\orcidlink{0000-0002-3048-5534}}\email[Mail should be sent to ]{afiorent@sissa.it}
\affiliation{%
 SISSA---Scuola Internazionale Superiore di Studi Avanzati, 34136 Trieste, Italy, European Union
}%
\author{Paolo Pegolo\,\orcidlink{0000-0003-1491-8229}}
\affiliation{%
 SISSA---Scuola Internazionale Superiore di Studi Avanzati, 34136 Trieste, Italy, European Union
}%
\author{Stefano Baroni\,\orcidlink{0000-0002-3508-6663}}
\affiliation{%
 SISSA---Scuola Internazionale Superiore di Studi Avanzati, 34136 Trieste, Italy, European Union
}%
\affiliation{%
 CNR-IOM---Istituto Officina Materiali, SISSA Unit, 34136 Trieste, Italy, European Union
}%
\date{\today}

\begin{abstract}
In the past few years, the theory of thermal transport in amorphous solids has been substantially extended beyond the Allen-Feldman model. The resulting formulation, based on the Green-Kubo linear response or the Wigner-transport equation, bridges this model for glasses with the traditional Boltzmann kinetic approach for crystals. The computational effort required by these methods usually scales as the cube of the number of atoms, thus severely limiting the size range of computationally affordable glass models. Leveraging hydrodynamic arguments, we show how this issue can be overcome through a simple formula to extrapolate a reliable estimate of the bulk thermal conductivity of glasses from finite models of moderate size. We showcase our findings for realistic models of paradigmatic glassy materials.
\end{abstract}

\maketitle

\section*{Introduction}

The theory of thermal transport in crystalline and disordered solids has witnessed a major advancement in the past few years~ \cite{marcolongo2015microscopic,isaeva2019modeling,simoncelli2019unified,simoncelli2022wigner,fiorentino2022green,caldarelli2022manybody,simoncelli2023thermal}. Until recently, the only way to simulate heat transport in glasses and complex crystals~\footnote{For complex crystal we mean crystals where anharmonic effects are strong with respect to the vibrational interband separations} was through the sheer application of the Green-Kubo (GK) linear-response theory, leveraging the heat-flux time series generated by Molecular Dynamics (MD) simulations~\cite{baroni2020heat, ercole2017accurate, bertossa2019theory, grasselli2021invariance, pegolo2022temperature}. Among the many merits of this approach is the full account of anharmonic effects of any order; however, the same cannot be said of quantum effects, which may be important at low temperatures, but cannot be captured by any classical method. The quest for a more convenient way of computing the thermal conductivity in glasses and complex crystals led to the unification of the approaches based on the Boltzmann Transport Equation (BTE) for crystalline materials~\cite{peierls1929kinetischen, peierls1955quantum}, and the Allen-Feldman (AF) model of thermal conduction in harmonic glasses~\cite{allen1989thermal, allen1993thermal}, resulting in the quasi-harmonic GK (QHGK) \cite{isaeva2019modeling} and Wigner-transport-equation (WTE)~\cite{simoncelli2019unified} approaches to heat conduction. The former method consists in an explicit evaluation of the GK expression of thermal conductivity from the anharmonic lattice dynamics of a solid close to mechanical equilibrium~\cite{isaeva2019modeling, barbalinardo2020efficient, fiorentino2022green}, while the latter exploits a different transport equation based on Wigner's quantum dynamics~\cite{simoncelli2019unified,simoncelli2022wigner, caldarelli2022manybody}. Despite their  differences, these two methods provide comparable results and are in fact conceptually equivalent, both being approximations of the same order to a general many-body approach based on GK linear-response theory and the Mori memory-function formalism~\cite{fiorentino2022green}. These methods suggest that the behavior of thermal conductivity in anharmonic solids can be understood in terms of the interplay between a quasi-particle kinetic mechanism, whereby traveling phonons propagate and scatter across many interatomic distances, and a diffusive regime, where disorder breaks the quasi-particle nature of phonons, and heat is transported through short-range hopping mechanisms.

These advances notwithstanding, the numerical simulation of heat transport in glasses remains a formidable task, because it requires the use of large finite models for the bulk systems, comprising up to several thousand atoms~\cite{feldman1993thermal, shenogin2009predicting, isaeva2019modeling, simoncelli2023thermal}. The same would be true even for crystalline materials, the difference being that, in such a case, lattice periodicity and the Bloch theorem can be exploited to map the simulation of a large model onto a number of computations performed for the vibrations of definite wavevector in the Brillouin zone (BZ), $\bm k$, of a unit cell comprising a much smaller number of atoms. These wavevectors are usually arranged in a regular grid whose number of nodes is the ratio between the total number of atoms in the bulk model and the number of atoms in the unit cell. The effective model that can thus be afforded has a linear dimension $L \sim 2\pi / k_{\mathrm{min}}$, $k_{\mathrm{min}}$ being the discretization step of the regular grid, and it may encompass up to several hundred thousand atoms.
Unfortunately, the spurious crystalline order introduced by the use of periodic boundary conditions (PBCs) in the simulation of finite glass models results in unphysical long-wavelength features in their transport properties~\cite{simoncelli2023thermal}, as it will be extensively illustrated in the following.

These issues call for a method able to efficiently and accurately extrapolate to infinite size the value of the thermal conductivity of aperiodic solids, such as glasses, without the need to artificially introduce nonphysical normal modes, which may lead to a gross overestimate of the final result. We devise such a method by leveraging arguments based on the hydrodynamics of solids, which exploit the natural partition of glassy normal modes into categories based on their localization in real space~\cite{allen1999diffusons}: \emph{propagons} are delocalized, low-frequency, vibrations whose typical wavelength is much larger than the interatomic distance and that therefore propagate almost freely as plane (sound) waves; \emph{diffusons} are intermediate-energy vibrations that spread diffusively; finally, \emph{locons} are localized excitations that populate the highest-frequency portion of the vibrational spectrum and they hardly contribute to the transport properties of the system.
Being long-wavelength vibrations, propagons are severely affected by the finite size of any glass model. In fact, the size of the system sets a lower bound to the minimum frequency accessible to the calculation, and determines the normal-mode spacing in the low-frequency region, which is already undersampled with respect to other portions of the spectrum, due to the vanishing of the Vibrational Density of States (VDOS) in the zero-frequency limit
~\cite{goldstone1961field,isaeva2019modeling, barbalinardo2020efficient}.
In this article we report on a method, which we dub \emph{hydrodynamic extrapolation}, able to lift these problems through a combination of two ingredients that can be inexpensively computed from finite models of moderate size: one is the QHGK contribution of diffusons and locons~\cite{isaeva2019modeling, barbalinardo2020efficient, fiorentino2022green}, the other being an effective low-frequency model for propagons. The term ``hydrodynamic'' is more often used in the context of the collective phonon transport in crystals~\cite{ghosh2022phonon}. Here, we adopt this term in a similar vein, as the underlying equations governing both crystals and glasses at low frequencies obey the same hydrodynamic principles~\cite{griffin1968brillouin}.

We underline that the hydrodynamic extrapolation applies above the so-called \textit{plateau} of the thermal conductivity as a function of temperature which is found in many glassy materials. The temperature dependence of $\kappa$ in glasses  features three universal characteristic trends~\cite{beltukov2013ioffe}: at very low temperatures, i.e. $T \lesssim 2\,\mathrm{K}$, the dominant scattering mechanism is the quantum tunneling between different local minima in the glass energy landscape, and $\kappa \sim T^2$~\cite{phillips1987two, buchenau1992interaction, lubchenko2003origin}; then, up to $\approx 30\,\mathrm{K}$, the thermal conductivity rises and saturates to a plateau value. Despite the absence of an established theoretical agreement in the literature, this phenomenon seems to be related to the crossover between the regime dominated by quantum processes and one where propagating waves are scattered by random disorder~\cite{buchenau1992interaction, lubchenko2003origin, schirmacher2006thermal, beltukov2013ioffe}. Above the plateau, the behavior of $\kappa$ is dictated by the anharmonic decay of nomal modes as prescribed by the QHGK theory. Since we do not have access to quantum-tunneling states that play a crucial role at low temperatures, in this work we focus on the third regime.

The structure of this Article is the following: first, we briefly review the QHGK method for computing the thermal conductivity of crystals and solids alike; then, we discuss the possibility of simulating heat transport in glasses from finite models in PBCs; we then introduce our hydrodynamic extrapolation technique, based on an effective low-frequency model for propagons; next, we illustrate our newly introduced method with a few numerical experiments performed for three paradigmatic classes of glass (amorphous silicon carbide, silica, and silicon); finally, we draw our conclusions. Following is a thorough description of the theoretical methods employed in our work.

\section*{Results and discussion}

\subsection*{The Quasi-Harmonic Green-Kubo Approximation} \label{Sec:Theory}

The quantum GK formula for the thermal conductivity tensor, $\kappa$, reads~\cite{green1952markoff,kubo1957statstical1,kubo1957statistical2}:

\begin{align} \label{eq:GK}
    \kappa = \frac{1}{VT} \int_0^\infty dt  \int_0^{\frac{1}{k_{\mathrm B}T}} \dd{\lambda} \expval{\widehat{J}(t-i\hbar\lambda)\widehat{J}(0)},
\end{align}
where $V$ is the system's volume, $k_{\mathrm B}$ is Boltzmann's constant, $T$ is the temperature, $\widehat J$ is any Cartesian component of the energy-flux operator in the Heisenberg representation, and isotropy is assumed throughout this paper, thus allowing us to dispose of Cartesian indices, unless strictly needed for clarity. Both crystalline and amorphous materials are normally described by finite-size models in PBCs, and normal modes can be labeled by wavevectors and band indices. Any Cartesian component of the harmonic energy flux operator reads~\cite{isaeva2019modeling, hardy1963energy}:
 
\begin{align}\label{eq:J-QHA}
    \begin{multlined}
        \widehat J = \hbar\sum_{\mathbf{q} \nu \nu'}\frac{\omega_{\mathbf{q} \nu}+\omega_{\mathbf{q} \nu'}}{2}v_{\mathbf{q}\nu\nu'}\had_{\mathbf{q}\nu} \ha_{\mathbf{q}\nu'} +\\ 
        +  \hbar\sum_{\mathbf{q} \nu \nu'}\frac{\omega_{\mathbf{q} \nu}-\omega_{\mathbf{q} \nu'}}{4}v_{\mathbf{q} \nu \nu'}  \bigl ( \ha_{-\mathbf{q}\nu}\ha_{\mathbf{q}\nu'}-\had_{\mathbf{q}\nu}\had_{-\mathbf{q}\nu'} \bigr ),
    \end{multlined}
\end{align}

where $(\mathbf{q},\nu)$ labels wavevector and band indices in the BZ, respectively; $\had_{\mathbf{q}\nu}$ and $\ha_{\mathbf{q}\nu}$ are the creation and annihilation operators, respectively; $\omega_{\mathbf{q} \nu}$ is the angular frequency of the normal mode;  $v_{\mathbf{q} \nu \nu'}$ is a generalized group velocity~\cite{isaeva2019modeling,caldarelli2022manybody} (see Eq.~\eqref{eq: generalized velocity}) which satisfies

\begin{align}
    v_{\mathbf{q}\nu\nu'}=v_{\mathbf{q}\nu'\nu}^*=-v_{-\mathbf{q}\nu'\nu}.
\end{align}
Here again, the Cartesian index of the velocity is dropped for notational simplicity. We stress that, while for crystals the classification of normal modes by their wavevectors reflects a physical symmetry of the systems, in glasses it is simply an artifact due to the use of PBCs. When large enough models are used to describe the glass, it is common practice to just sample the BZ center ($\Gamma, \mathbf{q}=0$), although ways of leveraging the phonon dispersions resulting from the use of PBCs in glasses have recently been proposed \cite{simoncelli2023thermal}. The first, number-conserving term, in Eq. \eqref{eq:J-QHA} is called \emph{resonant}, while the second, \emph{anti-resonant}, one gives negligible contributions in the small-linewidth limit and will be neglected in the followig. The diagonal term of the generalized velocity is the usual phonon group velocity: $v_{\mathbf{q}\nu\nu}=\nabla_{\mathbf{q}} \omega_{\mathbf{q} \nu}$. Eqs.~\eqref{eq:GK} and~\ref{eq:J-QHA} lead to the QHGK approximation to the thermal conductivity~\cite{isaeva2019modeling,barbalinardo2020efficient,fiorentino2022green}

\begin{align}
    \label{eq:k_markovian}
    \kappa = \frac{1}{V}\sum_{\mathbf{q} \nu \nu'} C_{\mathbf{q}\nu\nu'} v_{\mathbf{q}\nu\nu'} v_{\mathbf{q}\nu'\nu} \tau_{\mathbf{q}\nu\nu'},
\end{align}
where
 
\begin{align}
    C_{\mathbf{q}\nu\nu'} = \frac{\hbar^2 \omega_{\mathbf{q} \nu}\omega_{\mathbf{q} \nu'}}{T}\frac{n_{\mathbf{q} \nu}-n_{\mathbf{q} \nu'}}{\hbar(\omega_{\mathbf{q} \nu'}-\omega_{\mathbf{q}\nu})}
\end{align}
is the generalized two-mode heat capacity, ${n_{\mathbf{q} \nu}=(e^{\hbar\omega_{\mathbf{q} \nu}/k_{\mathrm B}T}-1)^{-1}}$ the Bose-Einstein distribution, and
 
\begin{align}\label{eq:tau qhgk}
    \tau_{\mathbf{q}\nu\nu'} = \frac{\gamma_{\mathbf{q} \nu}+\gamma_{\mathbf{q} \nu'}}{(\omega_{\mathbf{q}\nu}-\omega_{\mathbf{q} \nu'})^2+(\gamma_{\mathbf{q} \nu}+\gamma_{\mathbf{q} \nu'})^2}
\end{align}
is the generalized two-mode lifetime. In a nutshell, the QHGK approach entails two different---yet related---approximations, The first is the \emph{dressed bubble} approximation of four-point correlation functions among creation and annihilation operators; it amounts to neglecting vertex corrections to factorize four-point correlation functions into sums of products of two-point correlation functions~\cite{fiorentino2022green}. This means that normal modes decay independently of one another due to effective interactions with a common heat bath. The second approximation, often referred to as \emph{Markovian}, is to ignore any memory effects in the interaction between the heat bath and the normal modes~\cite{fiorentino2022green}. The combination of these two approximations in the QHGK approximation is that four-point correlation functions can be expressed in terms of single-body \emph{greater} Green's functions, given by ${g^>_{\bm{q}\nu}(t)=-i(n_{\bm{q}{\nu}}+1)\mathrm{e}^{-i\omega_{\bm{q}{\nu}}t -\gamma_{\bm{q}{\nu}}|t|}}$, where $\gamma_{\bm{q}{\nu}}$ is the linewidth of the normal mode labeled by $(\bm{q},\nu)$ as computed from Fermi's Golden Rule. The quasi-harmonic hypothesis requires $\gamma_{\bm{q}{\nu}}^2/\omega_{\bm{q}{\nu}}^2\ll 1$, implying that only almost-degenerate modes such that ${|\omega_{\bm{q}{\nu}}-\omega_{\bm{q}{\nu}'}|\lesssim \gamma_{\bm{q}{\nu}}+\gamma_{\bm{q}{\nu}'}}$ are allowed to contribute to the heat conductivity.

Glasses are, by definition, aperiodic. Yet, the effect of periodicity on large-enough models can be regarded as a size effect that vanishes in the thermodynamic limit. It is thus customary to treat glasses as finite models in a large supercell under PBCs sampled only at the zone center of the BZ. Omitting the sum over the (only) wave vector, $\Gamma$, and assuming the system is isotropic (thus, the thermal conductivity tensor is proportional to the identity matrix) the thermal conductivity reads:

\begin{align}\label{eq:kappa qhgk}
    \kappa = \frac{1}{3 V}\sum_{ \nu, \nu'}C_{\nu\nu'} |v_{\nu\nu'}|^2 \tau_{\nu\nu'}
\end{align}
The factor $1/3$ comes from the average over the three Cartesian directions, which are equivalent under the hypothesis of isotropy. At the $\Gamma$ point, the eigenvectors of the dynamical matrix can be chosen to be real, and $v_{\nu\nu'}$ becomes a real, antisymmetric matrix~\cite{isaeva2019modeling, barbalinardo2020efficient}. The harmonic limit is obtained when all the linewidths approach zero. In that case, the QHGK method reduces to the AF model~\cite{allen1989thermal, allen1993thermal}:

\begin{align}\label{eq:kappa_AF}
    \kappa &=\frac{\pi}{3V} \sum_{ \nu } C_\nu D_{\nu},
\end{align}
where $C_\nu$ is the modal specific heat and the modal diffusivity, $D_\nu$, is defined as 

\begin{align}
    D_\nu&=\sum_{\nu'} |v_{\nu\nu'}|^2\delta(\omega_\nu-\omega_{\nu'}).
\end{align}
The Dirac-delta function appearing in the diffusivity is due to the harmonic, limit ${\lim_{\gamma_\nu,\gamma_{\nu'} \to 0} \tau_{\nu\nu'}}$, of the Lorentzian generalized lifetimes defined in Eq.~\eqref{eq:tau qhgk}.

For crystalline materials, using a dense mesh in the BZ ($q$-mesh) allows one to efficiently increase the number of modes, and it is often required to get converged values of physical quantities, \emph{a fortiori} when the simulation cell is a (usually small) unit cell. In model glasses, the lowest frequency allowed in a cubic simulation cell with side $L$ is of order $2\pi c/L$, $c$ being the sound velocity; sampling a $q$-mesh mimics an artificial order by adding long-lived modes below this threshold, thus reflecting the periodicity induced by PBCs and determining a spurious increase of the thermal conductivity. Indeed, real glasses below this threshold frequency feature propagating waves whose wavelength is proportional to their inverse frequency, and whose decay depends on disorder~\cite{allen1999diffusons}. A replicated model glass cannot be disordered on a spatial scale larger than the original simulation cell. Thus, the spurious modes whose wavelength is larger than the simulation cell are unaffected by disorder and would propagate for longer times than the corresponding modes in an infinite, actually disordered, glass. In fact, finite wavevectors other than high-symmetry points at the BZ boundary have, in general, a finite group velocity, resulting in
 
\begin{align}\label{eq:kappa qhgk crystal}
    \begin{multlined}
    \kappa = + \frac{1}{3 V}\sum_{\mathbf{q} \nu}C_{\mathbf{q}\nu}v_{\mathbf{q}\nu} v_{\mathbf{q}\nu}\frac{1}{2\gamma_{q\nu}} + \\
    \qquad\qquad\qquad \frac{1}{3 V}\sum_{\mathbf{q}, \nu\neq\nu'}C_{\mathbf{q}\nu\nu'}v_{\mathbf{q}\nu\nu'} v_{\mathbf{q}\nu'\nu}\tau_{\mathbf{q}\nu\nu'}.
    \end{multlined}
\end{align}
The first term in the above equation is the same obtained by treating a crystal in the Relaxation Time Approximation (RTA) of the Boltzmann Transport Equation (BTE). When only third-order anharmonic effects are included, this term gives rise to the $1/T$ trend of $\kappa$ in crystals at high temperatures. Sampling the BZ of a glass model induces the same behavior, as recently observed in Ref.~\onlinecite{simoncelli2023thermal}, where a $3\times3\times3$ $q$-mesh is shown to determine a low-temperature divergence of the thermal conductivity of amorphous silica systems. The same does not happen for a genuinely disordered system with an equivalent number of atoms (i.e., $3^3$ times the number of atoms of the small model) sampled at the $\Gamma$ point; for such a system the thermal conductivity is a monotonically increasing function of temperature~\cite{simoncelli2023thermal}. Thus, replicating a small disordered system by sampling its reciprocal space is qualitatively different from simulating a correspondingly large system in real space, since in the former case disorder is suppressed at scales larger than its size. This behavior was also observed in GKMD calculations on glass samples replicated in real space~\cite{he2011heat,moon2018propagating}.

\subsection*{Hydrodynamic extrapolation of the thermal
conductivity}

The naive approach to deal with finite-size effects is to compute the heat conductivity for model systems with increasing size, $L$, up to convergence, $\kappa_\infty(T) = \lim_{L \to \infty} \kappa_L(T)$. 
The simplest way to extrapolate $\kappa_\infty(T)$ would then be to assume that $\kappa_L$ can be written as a power series in $1/L$ and perform a linear fit in $1/L$. This approach has two key issues: first, the whole procedure is computationally expensive, since it requires the preparation of samples with different sizes---at the very least three, in order to meaningfully fit a straight line, but in practice we observe that more points are usually needed; second, in cases where $L$ is too small for the linear contribution to be dominant, higher powers of $1/L$ might be necessary, thus increasing the number of points required to perform a reliable fit. In the following, we develop a method to bypass these obstacles making use of the features of the low-energy excitations naturally present in glasses.
\begin{figure}[t]
    \centering   
    \includegraphics[width=\columnwidth]{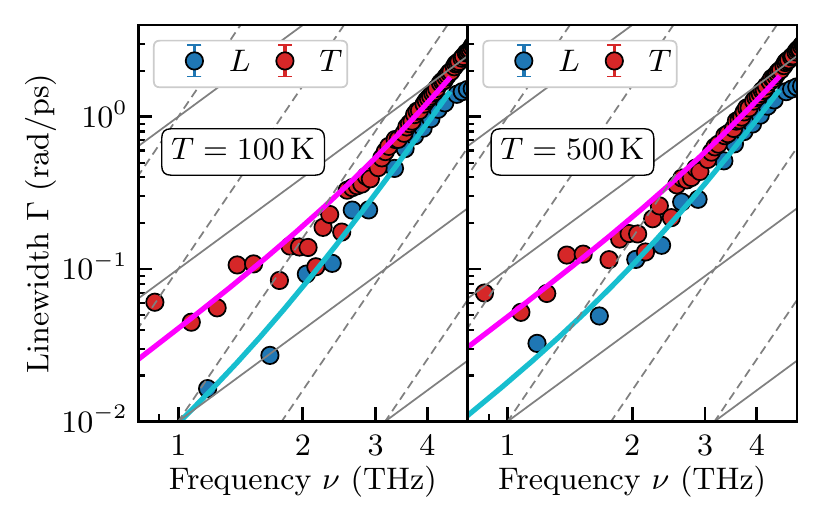}
    \caption{Longitudinal and transverse linewidths, $\Gamma$, computed from the DSF of an $13284$-atom model of $\asi$ at $100\,\mathrm{K}$ (left panel) and $500\,\mathrm{K}$ (right panel). The solid and dashed gray lines indicate $\sim \omega^2$ and $\sim \omega^4$ behaviors, respectively.}
    \label{fig:Si_Gamma}
\end{figure}

As mentioned above, propagons, diffusons, and locons differ by the degree of localization they feature. This can be observed in the Dynamical Structure Factor (DSF) that, for a harmonic system, is defined as~\cite{feldman1999numerical, moon2018propagating, larkin2014thermal}:

\begin{align}\label{eq:dsf harmonic}
    S_b^0(\omega,\mathbf{Q})=\sum_{\nu}\delta(\omega-\omega_\nu)|\langle \nu| \mathbf{Q},b\rangle|^2    
\end{align}
where the scalar product between normal modes and plane-wave states is
 
\begin{align}
    \langle \nu| \mathbf{Q},b\rangle=\frac{1}{\sqrt{N_{\mathbf{Q}}}}\sum_{i\alpha} \varepsilon_\alpha^b (\mathbf{Q})\epsilon_{i\alpha}^\nu e^{i \mathbf{Q}\cdot \mathbf{R}_I},
\end{align}
$\alpha$ being the Cartesian component; ${\mathbf{Q} = \frac{2\pi}{L}(n,m,l)}$, with $n,m,l \in \mathbb{N}$, is a wavevector in a cubic supercell of side $L$; $\mathbf{R}_I$ the position of the $I$th atom; $\bm{\epsilon}^\nu$ the eigenvector of the $\nu$th normal mode; and $\bm{\varepsilon}^b(\mathbf{Q})$ a polarization (unit) vector (see \emph{Methods}). The latter can be chosen to be parallel or perpendicular to $\mathbf{Q}$ and can be labeled by $b=L,T_1,T_2$, the longitudinal (parallel to $\mathbf{Q}$) and transverse (perpendicular) branches, respectively. Transverse branches are degenerate, so we indicate with $T$ the contributions of both transverse branches.
The DSF can be generalized to take into account weak anharmonic effects as
 
\begin{align}\label{eq:dsf anharmonic}
    S_b(\omega,\mathbf{Q})=\sum_{\nu}\frac{1}{\pi}\frac{\gamma_\nu}{\gamma_\nu^2+(\omega-\omega_\nu)^2}|\langle \nu| \mathbf{Q},b\rangle|^2,
\end{align}
where the Dirac-delta in Eq.~\eqref{eq:dsf harmonic} is replaced by a Lorentzian function centered on the mode's angular frequency, whose spread is given by the mode's linewidth. Eq.~\eqref{eq:dsf anharmonic} gains a temperature dependence through the anharmonic linewidths with respect to Eq.~\eqref{eq:dsf harmonic}.
The low-frequency, small-wavevector, portion of each branch of the DSF features an almost linear dispersion, ${\omega \simeq c Q}$, typical of acoustic phonons. In other words, $S_b(\mathbf{Q}, \omega)$ is a peaked function centered at $\omega_Q=c_b Q$, $c_{T,L}$ being the transverse/longitudinal speed of sound. For low-enough $Q$, the profile of $S_b(\omega,\mathbf{Q})$ can be fitted as a function of the angular frequency with a Lorentzian profile,

\begin{align}\label{eq:dsf anharmonic-2}
    S_b(\omega,\mathbf{Q}) \approx \frac{A_b(\mathbf Q)}{\pi} \frac{\Gamma_b(\mathbf Q)}{(\omega-c_b Q)^2+\Gamma_b(\mathbf Q)^2},
\end{align}
allowing one to evaluate the speed of sound as well as the wavevector dependence of the effective width, $\Gamma_b(\mathbf{Q})$. Under the assumption of isotropy, $S_b(\omega,\mathbf{Q})=S_b(\omega,Q)$ and, consequently, $\Gamma_b(\mathbf{Q})=\Gamma_b(Q)$.
Propagons are identified as those low-frequency, low-wavevector, normal modes with linear dispersion that populate the first portion of the DSF. 
The increasing broadening of the dispersion identifies a cutoff frequency for propagons, $\omega_\mathrm{P}$, referred to as the Ioffe-Regel limit~\cite{ioffe1960progress}.

The peaked shape of $S_b(\omega,Q)$ suggests the $|\mathbf{Q},b\rangle$s are long-lived excitations. In the long wavelength limit, they form an almost-orthonormal basis even for aperiodic materials such as glasses~\cite{feldman1999numerical, suppmat}, whose time-evolution for positive times---as inferred from the DSF---is proportional to $e^{-i \bigl (c_bQ-i\Gamma_b(Q) \bigr )t}$. Unlike the case of normal modes, which are eigenvectors of the harmonic (disordered) Hamiltonian, plane-wave states decay both through anharmonic interactions and as an effect of harmonic disorder. In fact, even the harmonic DSF, $S_b^0$, features a finite width for any $Q$~\cite{feldman1999numerical} without the need for any anharmonic effect. This does not happen in a periodic system. 
In order to extrapolate the thermal conductivity of larger systems, we separate the contribution of propagons from that of from diffusons and locons:

\begin{align}\label{eq:sep_kappa_prop_and_diff}
    \kappa=\kappa_\mathrm{P}+\kappa_\mathrm{D}.
\end{align}
The first contribution, due to propagons, involves only normal modes below the cutoff frequency, $\omega_\mathrm{P}$. The second contribution involves normal modes above such frequency, diffusons and locons, and it is mainly due to diffusons.
The propagon contribution can be approximately written on the basis of acoustic plane waves $|\mathbf{Q},b\rangle$~\cite{feldman1999numerical,taraskin2000propagation}. The overall idea is to switch from the basis of normal modes, whose decay can only be anharmonic, to a basis of delocalized modes whose decay, encompassed in the definition of $\Gamma_b(Q)$, entails a combination of anharmonicity and disorder. We demonstrate this in the Methods section, while giving here just a sketch of the proof. 

Let us consider the energy flux of Eq.~\eqref{eq:J-QHA}; the resonant propagon contribution to the energy flux reads:

\begin{align}
    \widehat J = \frac{1}{V}\sum_{\mathbf{Q} b b'} J_{\mathbf{Q}}^{bb'}\had_{\mathbf{Q} b} \ha_{\mathbf{Q} b'},
\end{align}
where
 
\begin{align}
    \ha_{\mathbf{Q} b}=\sum_{\nu} \langle \nu| \mathbf{Q},b\rangle \ha_{\nu},    
\end{align}
and $J_{\mathbf{Q}}^{bb'}$ is the matrix element of $\widehat{J}$ in the plane-wave basis. Similarly to what happens in crystals, such matrix elements can be shown to have a block-matrix structure in the wavevector indices

\begin{align}
    J_{\mathbf{Q} \mathbf{Q}'}^{bb'} \approx \delta_{\mathbf{Q}\mathbf{Q}'}J_{\mathbf{Q}}^{bb'}.
\end{align}
The two-point correlation function of the acoustic waves is related to the anharmonic DSF by
 
\begin{align}
    \langle \ha_{\mathbf{Q} b}(t) \had_{\mathbf{Q} b'}\rangle=\sum_{\nu}|\langle \nu| \mathbf{Q},b\rangle|^2 \langle \ha_\nu\had _\nu\rangle e^{-i\omega_\nu t-\gamma_\nu|t|},
\end{align}
that can be rewritten as
 
\begin{align}\label{eq:correlation-propagons}
    \begin{split}
        \langle \ha_{\mathbf{Q} b}(t) \had_{\mathbf{Q} b'}\rangle &\approx \delta_{bb'}\frac{1}{2\pi}\int e^{-i\omega t}S_b(\omega,\mathbf{Q}) (n(\omega,T)+1) \dd{\omega}\\
        &\approx \delta_{bb'}(n(c_bQ,T)+1)e^{-ic_bQ t-\Gamma_b(Q) |t|}.
    \end{split}
\end{align} 
The propagon contribution to Eq.~\eqref{eq:GK} can thus be evaluated from Eq.~\eqref{eq:correlation-propagons}, resulting in
 
\begin{align}\label{eq:kappa propagons}
    \kappa_\mathrm{P}&=\frac{1}{3V}\sum_{\mathbf{Q}, b} C(c_b Q)c_b^2\tau_{Q b}.
\end{align}
The last equation defines the acoustic excitations' lifetimes, $\tau_{Q b}=\frac{1}{2\Gamma_{b}(Q)}$, and the frequency-dependent heat capacity, $C(\omega)=\hbar \omega \frac{\partial n_\omega}{\partial T}$. In principle, an inter-band contribution should also appear between the transverse and longitudinal branches, with a Lorentzian weight as in Eq.~\eqref{eq:k_markovian}. However, longitudinal and transverse branches are energetically well separated, i.e., ${|(c_\mathrm{L}-c_\mathrm{T})Q|\gg \Gamma_\mathrm{L}(Q)+\Gamma_\mathrm{T}(Q)}$, both because in $3$D materials ${c_\mathrm{L}>\sqrt{4/3} c_\mathrm{T}}$~\cite{landau1960theory}, and because propagons are, by definition, modes with a sharp linear dispersion relation, i.e., $c_bQ \gg \Gamma_b(Q)$. 
Eq.~\eqref{eq:kappa propagons} is reminiscent of the RTA for crystals, and a similar equation appeared many times in the literature on harmonic glasses (e.g., in Refs.~\onlinecite{feldman1999numerical, larkin2014thermal, cahill1994thermal}), since it is just a statement that, at large-enough length scales, glasses can be effectively described by a continuous elastic model. Here, this fact is derived from first principles (i.e., from the QHGK theory) also including anharmonic effects.

The final step to evaluate the infinite-size limit of $\kappa$ is to map Eq.~\eqref{eq:kappa propagons} onto the kinetic theory of gases. The sum over plane waves can be recast as an integral over angular frequency in the ${L\to\infty}$ limit using the fact that, for propagons, the linear dispersion relation, $\omega=c Q$, is valid, and introducing per-branch densities of states, ${\rho_b(\omega)=\frac{1}{V} \sum_{\mathbf{Q}} \delta(\omega-c_{b}Q)}$. This yields:

\begin{align}\label{eq:kappa_propagons_continuous}
    \kappa_\mathrm{P} = \sum_{b=L,T} \frac{c_b^2}{3} \int_0^{\omega_\mathrm{P}} C(\omega) \rho_b(\omega) \frac{1}{2\Gamma_b(\omega)} \dd \omega.
\end{align}
Acoustic excitations feature a density of states of the Debye form, i.e., $\rho_\mathrm{T}(\omega)=\frac{\omega^2}{\pi^2 c_\mathrm{T}^3}$ and $\rho_\mathrm{L}(\omega)=\frac{\omega^2}{2\pi^2 c_\mathrm{L}^3}$. Once $\kappa_\mathrm{P}$ is available, one computes the hydrodynamic extrapolation of the thermal conductivity as 

\begin{multline}\label{eq:kappa_hydro_final}
    \kappa_{\mathrm{hydro}} = \kappa_\mathrm{P} \\ + \frac{1}{3V} \sum_{\nu,\nu'} \Theta(\omega_{\nu,\nu'}-\omega_\mathrm{P}) C_{\nu\nu'} v_{\nu\nu'} v_{\nu'\nu} \tau_{\nu\nu'},
\end{multline}
where the second term comes from Eq.~\eqref{eq:kappa qhgk}, and the Heaviside-theta limits the sum to the nonpropagonic part of the spectrum. 

The choice of $\omega_\mathrm{P}$ must satisfy certain requirements. For the model to be valid, all the normal modes below $\omega_\mathrm{P}$ should be \emph{bona fide} propagons for both polarizations, so as to avoid the mixed transverse-diffusons/longitudinal-propagons regime. Thus, even though the IR limit is different for the two polarizations, $\omega_\mathrm{P}$ should be less than the smallest of the two, which is generally the transverse one. Slightly lower  values can also be preferred to guarantee that all the hypotheses regarding linear dispersion are valid. The SI includes a dedicated section that provides practical suggestions on selecting the appropriate value of $\omega_\mathrm{P}$. Another important point is how to get a good estimate of $\Gamma_b$ as a function of $\omega$, which is needed to compute $\kappa_\mathrm{P}$ from Eq.~\eqref{eq:kappa_propagons_continuous}. This is done by first fitting the computed DSFs, Eq. \eqref{eq:dsf anharmonic}, as a function of $\omega$, at fixed $Q$, to a Lorentzian function, Eq. \eqref{eq:dsf anharmonic-2}. We then use the linear dispersion, $\omega_b = c_b Q$, to get the functional dependence of of the linewidth upon frequency,  $\Gamma_b(\omega)$. We finally fit this dependence to a quartic polynomial,
$\Gamma_b(\omega) = a \omega^2 + b \omega^4$ with $a,b>0$, as it can be observed in Fig.~\ref{fig:Si_Gamma} for the linewidths of $\asi$. A model of this form was proposed also in Ref.~\onlinecite{braun2016size}, where the quadratic and quartic terms come, respectively, from the Umklapp and isotopic scattering of long-wavelength phonons in crystalline silicon. More generally, the $\Gamma \sim \omega^2$ trend is required by hydrodynamics~\cite{kadanoff1963hydrodynamic,griffin1968brillouin}, while the $\Gamma \sim \omega^4$ behavior---also found in experiments~\cite{baldi2011elastic}---can be rationalized in the continuous limit through random media theory~\cite{izzo2018mixing,izzo2020rayleigh}, or for atomistic models through harmonic perturbation theory, such as in the case of crystals with mass disorder~\cite{garg2011role,mahan2019effect} or random spring constants~\cite{allen1998evolution}. The crossover between the $\sim\omega^2$ and $\sim\omega^4$ behaviors in the sound attenuation can be understood analyzing nonaffine displacements in amorphous solids~\cite{baggioli2022theory}. A rule of thumb to assess whether the bandwidth can be meaningfully computed from a glass sample of given size is to check whether the minimum frequency available at that size, $\omega_\mathrm{min}$, is smaller than the crossover frequency between anharmonicity and disorder; namely, $\omega_\mathrm{min} \lesssim \sqrt{a/b}$.

\begin{figure*}[t]
    \centering
    \includegraphics[width=1.6\columnwidth]{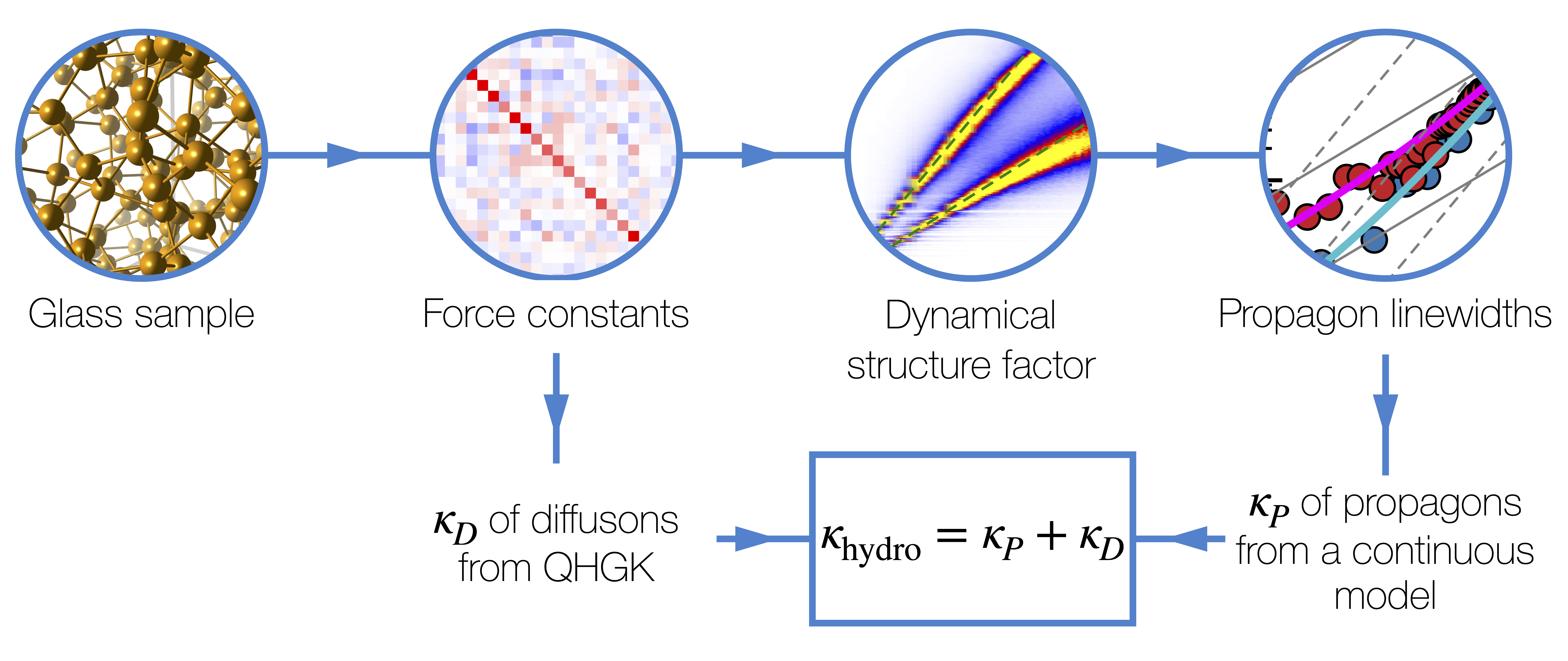}
    \caption{Graphical summary of the workflow necessary to implement the hydrodynamic extrapolation of the heat conductivity described in this work.}
    \label{fig:scheme}
\end{figure*}
\begin{figure}[t]
    \centering
    \includegraphics[width=\columnwidth]{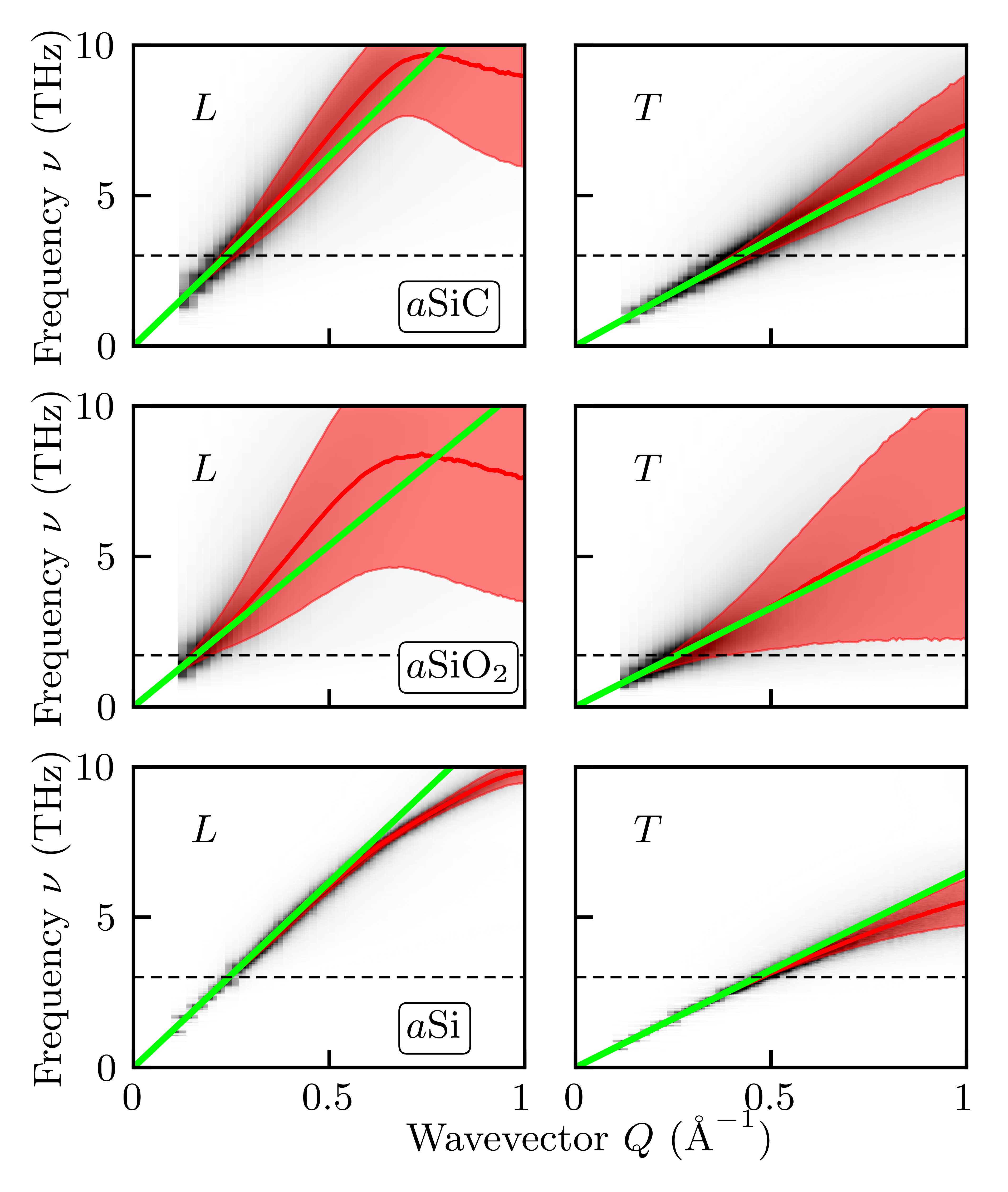}
    \caption{Anharmonic DSFs of three amorphous solids at $300\,\mathrm{K}$. For each material, the two polarizations, longitudinal ($L$) and transverse ($T$), are reported in different subplots. The colormaps represent the intensity of the DSF computed from Eq.~\eqref{eq:dsf anharmonic}. The red, solid, line and its spread are mode-frequencies and linewidths obtained from Lorentzian fits of $S_b(\omega, \mathbf{Q})$ for fixed values of $\mathbf{Q}$, as explained in the Methods. The green, straight lines are linear fits of the low-frequency portions of the DSFs' mode-frequencies. The black, dashed, lines are the values of $\omega_\mathrm{P}/2\pi$ for the three materials.}
    \label{fig:dsf}
\end{figure}
In order to validate Eq.~\eqref{eq:kappa_hydro_final}, we compare its predictions with those of an improved version of the naive procedure outlined at the beginning of this section. Namely, instead of a polynomial fit of $\kappa_L(T)$ in $1/L$, which can easily incur overfitting due to the small number of data points usually at one's disposal, we opt to use a different functional form suggested by the continuous model we are using:
 
\begin{align}\label{eq:kappa fit size}
    \kappa(L,T)=-A\arctan(B/L)+\kappa_\infty,
\end{align}
where $A$ and $B$ are positive parameters. This ansatz can be derived from Eq.~\eqref{eq:kappa_propagons_continuous} under the following assumptions: \textit{i)} that the classical limit is valid [i.e., $C(\omega)=k_{\mathrm B})$]; \textit{ii)} that the form of the linewidth is $\Gamma(\omega)=a\omega^2+b\omega^4$; and \textit{iii)} that the only size effect comes from the value of the minimum frequency available at each size, $\omega_{\mathrm{min}}= 2 \pi c/L$. In principle, one would need two sets of parameters, one for each polarization, but this would unnecessarily complicate the fitting procedure. Using just one set of parameters, as in Eq.~\eqref{eq:kappa fit size}, amounts to saying that we are effectively averaging the contributions due to each polarization. 
Under these simplifying hypotheses, Eqs.~\eqref{eq:kappa_propagons_continuous} and~\eqref{eq:kappa_hydro_final} become:
 
\begin{align}
    \begin{split}
        \kappa(L,T) &= \frac{k_{\mathrm B}}{2\pi^2 c} \int_{2\pi c/L}^{\omega_\mathrm{P}} \omega^2 \frac{1}{2(a\omega^2+b\omega^4)} \dd \omega+\kappa_D\\
        &=\kappa_\infty - \frac{k_{\mathrm B}}{4\pi^2 c \sqrt{ab}}\arctan\left ( \frac{2\pi c}{L} \sqrt{\frac{b}{a}} \right ). 
    \end{split}
\end{align}
Since only low-energy modes with $0 \le Q \le 2\pi c/L$ are  involved, the classical approximation of the modal specific heat is further justified. This ansatz not only reproduces the $1/L$ trend for large-enough sizes, but it also captures the non-linearity (in particular the convexity) of our data. Fig.~\ref{fig:scheme} summarizes how the hydrodynamic extrapolation scheme outlined above is applied to estimating the bulk thermal conductivity of glass samples.

\subsection*{Numerical experiments}

\begin{figure*}[t]
    \centering
    \includegraphics[width=2\columnwidth]{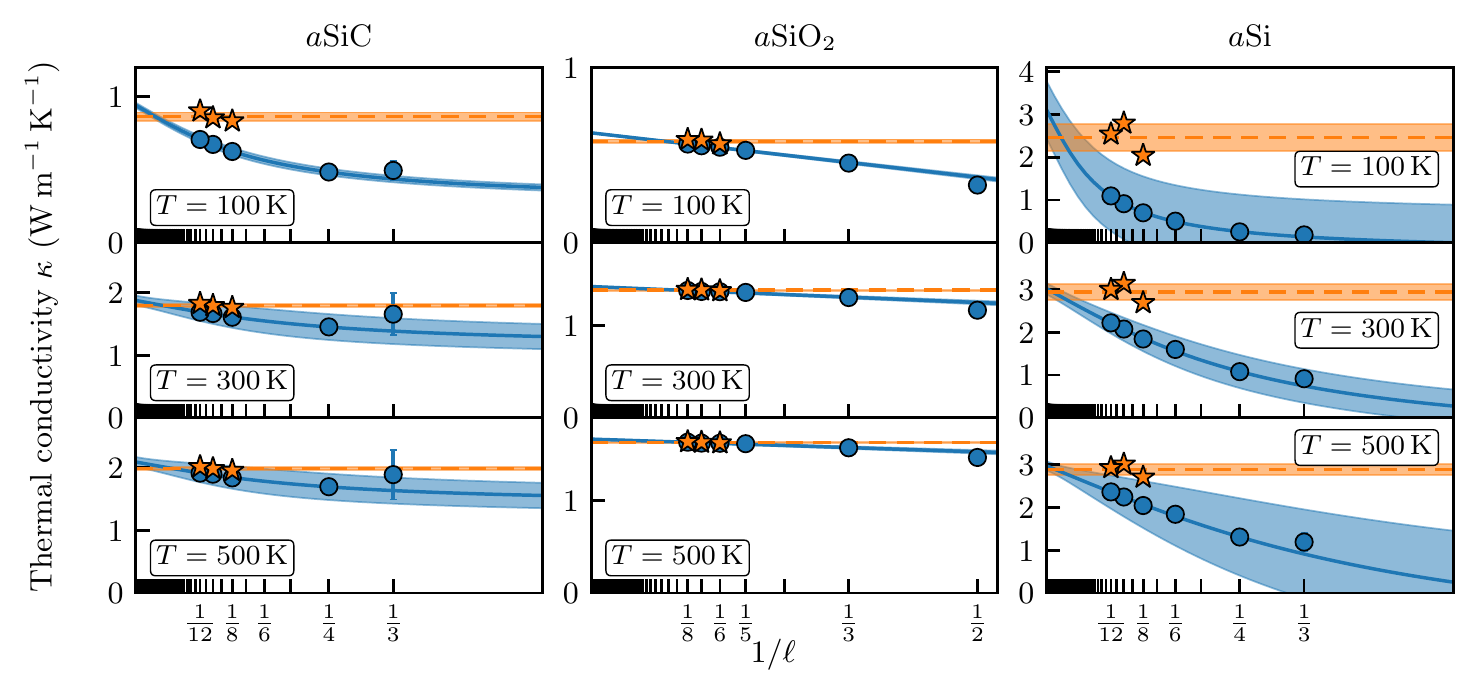}
    \caption{Size scaling of the thermal conductivity for three glassy materials at different temperatures. Light blue circles are QHGK calculations performed on samples of different sizes. Error bars are standard deviations over up to ten equivalent configurations. Light blue solid lines and shaded areas are fits done according to Eq.~\eqref{eq:kappa fit size} and respective error bars. Orange stars are hydrodynamic extrapolations of $\kappa$ done on top of QHGK calculations. Orange dashed lines denote the average value of $\kappa_\mathrm{hydro}$ over different sizes, while the shaded areas represent the respective standard deviation.}
    \label{fig:kappa extrapolation}
\end{figure*}

We now showcase the ability of our method to accurately predict the bulk limit of the thermal conductivity for three paradigmatic glasses, chosen because they have been extensively investigated in the literature and because they are representative of different regimes as concerns the role of propagating modes: amorphous silica ($\asio$) is a material whose thermal transport properties are mainly determined by diffusons; in amorphous silicon ($\asi$) transport is dominated by propagons; the situation is somewhat intermediate in amorphous silicon carbide ($\asic$)~\cite{larkin2014thermal, moon2019thermal}. 

For each material, the heat conductivity is averaged over ten different samples obtained through a melt-and-quench procedure described in the \emph{Methods} section. 

Second- and third-order interatomic force constants as well as normal-mode frequencies and lifetimes have been computed with the $\kappa$ALD$o$~\cite{barbalinardo2020efficient} code, using forces obtained by the LAMMPS MD code~\cite{LAMMPS}.

For notational convenience, the size-scaling data reported in the following are plotted against the number replicas of the fundamental simulation cell repeated along each Cartesian component, $\ell=L/L_0$, where $L$ is the actual linear size of the simulation cell being employed and $L_0$ is the size of the smallest cell used for each material, which contains 8, 24, and 8 atoms for $\asic$, $\asio$, and $\asi$, respectively.

\subsubsection*{Amorphous silicon carbide}

The top panel of Fig.~\ref{fig:dsf} shows the DSF of $\asic$. 
The dispersion for the longitudinal branch is linear up to approximately $8.5\,\mathrm{THz}$, in agreement with the IR limit found in Ref.~\onlinecite{moon2019thermal}. From the slope of the linear dispersion, the speeds of sounds are computed to be $c_\mathrm{L}\approx 8.0\,\mathrm{km~s^{-1}}$ and $c_\mathrm{T}\approx 4.4 \,\mathrm{km~s^{-1}}$.
We choose a cutoff $\omega_\mathrm{P}/2\pi = 3 \,\mathrm{THz}$, in order to stay well within the linear dispersion regime for both polarizations. In the left panel of Fig.~\ref{fig:kappa extrapolation}, we report the resulting $\kappa_{\mathrm{hydro}}$ for samples ranging from $4096$ atoms ($\ell=8$) to $13824$ atoms ($\ell=12$). Size effects are quite important at $100\,\mathrm{K}$, where the bulk conductivity is almost $35\%$ larger than one of our largest models. The hydrodynamic extrapolation performs well with relatively small samples. At higher temperatures, size effects are less striking; nonetheless, the hydrodynamic extrapolation still improves results with respect to plain QHGK calculations. 

\subsubsection*{Amorphous silica}

The method is tested also on amorphous silica, with a $\omega_\mathrm{P}/2\pi = 1.7 \,\mathrm{THz}$, which is quite low. The reason for this is that, as shown in the middle panel of Fig.~\ref{fig:dsf}, the IR limit for longitudinal waves is reached for frequencies as small as $\approx 2 \,\mathrm{THz}$, which is quite smaller than those of amorphous silicon and silicon carbide, whose values are $\approx 8.5 \,\mathrm{THz}$ and $\approx 9 \,\mathrm{THz}$, respectively~\cite{moon2019thermal}. This fact suggests that the contribution of propagons to the thermal conductivity of $\asio$ is minor, and that size effects are rather small. Our calculations confirm these statements: the central panel of Fig.~\ref{fig:kappa extrapolation} shows that not only is the QHGK thermal conductivity of the $8000$-atoms model already large enough to be practically at convergence in size, but that its dependence on inverse size, $1/\ell$, is practically linear for any meaningful smaller size: thus, in this case, there is not much to gain in computing hydrodynamic corrections for $\asio$, if compared with other materials.

\subsubsection*{Amorphous silicon}

$\asi$ features a sharp linear dispersion in the low-frequency region where the propagon linewidth is considerably smaller than in $\asic$ and $\asio$, as one can see from the bottom panel of Fig.~\ref{fig:dsf}. This reflects on the large value of the propagonic contribution to the thermal conductivity: in fact, with ${\omega_\mathrm{P}/2\pi=3\,\mathrm{THz}}$, $\kappa_\mathrm{P}$ constitutes about $60\%$ of the thermal conductivity at room temperature, a fraction that inevitably grows for lower temperatures, due to the increasing importance of low-energy modes. The consequence of the dominance of the propagons in the value of $\kappa$ is that size effects become rather crucial in $\asi$. In particular, the thermal conductivity has a conspicuous nonlinear dependence on $1/\ell$ in the size range accessible to numerical simulations. Since such nonlinearity is mainly due to propagons, it is well captured by the fitting procedure discussed above.

In order to verify our results are not due to overfitting, we performed some classical GKMD at high temperatures on systems with larger sizes, up to $64000$ atoms ($\ell=20$), which are not affordable through lattice methods such as QHGK. As reported in the Supplementary Information (SI)~\cite{suppmat}, even with $64000$ atoms $\kappa$ is still far from convergence; however, the computed GKMD thermal conductivities are compatible with the fit on the QHGK data, thus independently confirming the validity of the whole procedure.  
Notwithstanding, the hydrodynamics extrapolation on samples larger or equal than $\ell=8$ ($N=4096$) leads to bulk thermal conductivity values compatible with the fitted ones, as shown in the right panel of Fig.~\ref{fig:kappa extrapolation}.

\subsection*{Comparison with existing methods} \label{sec:comparison rwte}

A different approach has been recently proposed to extrapolate the value of the bulk thermal conductivity in glasses from simulations performed for small models and leveraging the symmetry properties of the fictitious crystal resulting from the periodic replica of the simulation cell in PBCs  \cite{simoncelli2023thermal}. The proposed method consists of two steps: a finite model for a glass under PBCs is first treated as a genuine crystal, i.e., its vibrational properties are sampled on a dense $q$-mesh in the BZ: we call this procedure a \emph{crystalline model} for the glass. The $1/T$ divergence of the heat conductivity, resulting from crystalline order, is then avoided by convolving the Lorentzian lineshape of the unphysical low-frequency modes introduced by the periodic replicas of the fundamental simulation cell with a smearing function whose width, $\eta$, is carefully chosen as described below. While choosing the smearing function to be Lorentzian would result in a Lorentzian smeared lineshape whose width would be the sum of the original and smearing linewidths, it was found that using a Gaussian smearing function would lead to a better numerical behavior of the procedure. Of course, the heat conductivities computed with this method depend on the smearing width, and they would in fact diverge, in the low-temperature limit, for vanishing widths. In Ref. \onlinecite{simoncelli2023thermal} the choice of the smearing width relied on the existence of a plateau in the dependence of the heat conductivity on it. When such a plateau exists, the width so chosen should be a compromise between a large enough value allowing to encompass several normal modes and small values necessary not to mimic spuriously large scattering sources.

In a nutshell, this protocol forces the QHGK (or, equivalently, WTE) method for a nonphysical crystal---whose unit cell is the whole simulation representing the glass model---to behave like the AF model for low-frequency modes whose linewidths are smaller than the smearing width---i.e., the modes responsible for the unphysical overestimate of $\kappa$---while doing almost nothing to modes with higher frequency/larger linewidth.

This method was applied to amorphous silica ($\asio$)~\cite{simoncelli2023thermal}. A computationally affordable $\asio$ crystalline model treated with a $3 \times 3 \times 3$ $q$-mesh, once regularized with the above procedure, yielded values of $\kappa$ compatible with the temperature-dependent thermal conductivity of a genuinely larger system with an equivalent number of atoms, thus validating the whole procedure for $\asio$. While we are able to reproduce these results and find a plateau in the dependence of $\kappa$ on the smearing width for our $\asio$ samples, we cannot do so for other materials, where no plateau is found, thus making it difficult in general to fix the value of a suitable smearing width. In the SI~\cite{suppmat} we test the regularization procedure on amorphous silica, amorphous silicon carbide, and amorphous silicon.

We rationalize these difficulties as follows. The low-temperature behavior of the heat conductivity of bulk solids is brought about by the interplay of anharmonicity and disorder. The former gives rise to a $\omega^2$ dependence of $\Gamma_b$ at low frequencies and dominates the lowest-frequency portion of the propagon spectrum, while the effects of the latter, contributing as $\omega^4$, emerge later and dominate at larger frequencies. Sampling a periodically repeated cell of linear size $L$ on regular $q$-mesh completely misses the effect of disorder for the modes introduced by periodicity of frequency smaller than ${\sim 2\pi c/ L}$. The convolution of the vibrational lineshapes with a smearing function mimics the introduction of an effective boundary-scattering term in the form of a constant linewidth, rather than the correct frequency-dependent scattering due to disorder. The net effect is that the thermal conductivity is forced to quickly go to zero for vanishing temperatures. For materials where the contribution of propagons to the heat conductivity is small, such as $\asio$~\cite{larkin2014thermal}, the overdamping of low-frequency modes does not constitute a problem; vice versa, it fails to capture the main damping mechanisms that abate the heat conductivity in the low-temperature regime in materials where the propagons' contribution is significant, if not dominant. 
More details on calculations based on Ref.~\onlinecite{simoncelli2023thermal} can be found in the SI~\cite{suppmat}.

\subsection*{Temperature dependence of thermal conductivity}

\begin{figure}
    \centering
    \includegraphics{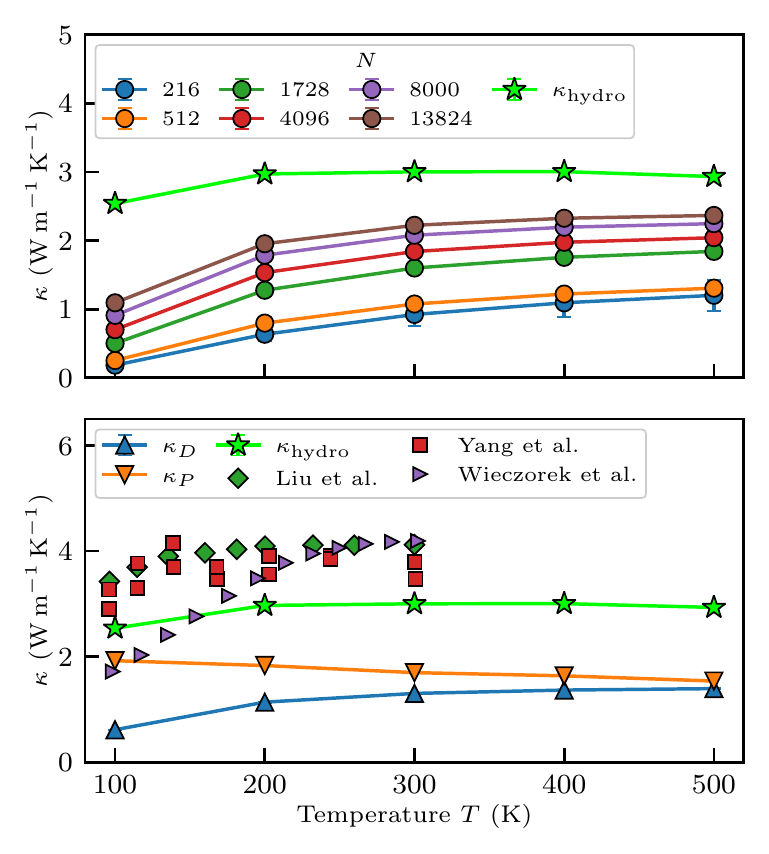}
    \caption{Thermal conductivity of $\asi$ as a function of temperature. The upper panel compares the hydrodynamic extrapolation of the thermal conductivity and QHGK calculations. The lower panel provides a breakdown of $\kappa_{\mathrm{hydro}}$ into its the propagon contribution, $\kappa_\mathrm{P}$, and the diffuson contribution, $\kappa_D$; the results of the calculations are compared with experimental measurements found in the literature.}
    \label{fig:kappa vs T}
\end{figure}

It is instructive to analyze the temperature dependence of $\asi$ since it is the material, among those at our disposal, where propagons contribute the most to the value of the thermal conductivity. The upper panel of Fig.~\ref{fig:kappa vs T} compares the hydrodynamic extrapolation with the QHGK calculations performed on progressively larger systems. The lower panel provides a breakdown of the extrapolated $\kappa$ into the propagon contribution, $\kappa_\mathrm{P}$, and the diffuson one, $\kappa_D$. The contribution of diffusons resembles the one of a standard AF calculation~\cite{feldman1999numerical,allen1999diffusons}, ranging from small values at low temperatures to some constant value around room temperature. Conversely, the behavior of $\kappa_\mathrm{P}$ is rather surprising, as it decreases with increasing temperature. This trend cannot be explained by a purely harmonic theory, where the temperature dependence is solely determined by the heat capacity~\cite{allen1993thermal, isaeva2019modeling}, and it is thus an anharmonic effect. 


The comparison between the hydrodynamic extrapolation and QHGK calculations highlights the significance of the former, especially at lower temperatures, as illustrated in both the upper panel of Fig.~\ref{fig:kappa vs T} and Fig.~\ref{fig:kappa extrapolation}. This observation can be attributed to two key factors: (\emph{i}) at low temperatures, only propagons are populated and contribute significantly to thermal conductivity. Consequently, it becomes essential to employ a continuous model that accounts for frequencies below the minimum frequency allowed by the finite-size sample. (\emph{ii}) Numerically accurate QHGK results require the frequency spacing between modes to be smaller or comparable to the anharmonic linewidths, as specified in Eq.~\eqref{eq:kappa qhgk}. As the anharmonic linewidths decrease with temperature, achieving meaningful QHGK thermal conductivity values at low temperatures necessitates a denser VDOS, i.e., larger systems.

The lower panel of Fig.~\ref{fig:kappa vs T} also includes experimental measurement of the thermal conductivity of $\asi$ on samples of relatively large sizes obtained with chemical vapor deposition techniques. The data are sourced from Liu et al.~\cite{liu2009high}, Yang et al.\cite{yang2010anomalously}, and Wieczorek et al.\cite{wieczorek1989}. The respective films have thicknesses of $80\,\mathrm{\mu m}$, $1.6$-$2.8\,\mathrm{\mu m}$, and  $2$-$3.6\,\mathrm{\mu m}$. The motivation behind selecting these larger samples from the literature is to minimize size effects and obtain an estimate of the bulk thermal conductivity of $\asi$.

The experimental values of $\kappa$ exhibit good agreement among themselves above approximately $200\,\mathrm{K}$. However, at lower temperatures, Wieczorek et al.~\cite{wieczorek1989} reports slightly smaller values compared to the other sources. Overall, the behavior of the thermal conductivity is well captured by $\kappa_\mathrm{hydro}$, including the relatively high value observed around $100\,\mathrm{K}$. Nevertheless, our model appears to slightly underestimate the thermal conductivity across the entire temperature range. We identify two primary reasons for this discrepancy.

One factor is the choice of the force field, specifically the Tersoff force field, which is not explicitly designed to accurately reproduce thermal conductivity. As a result, achieving numerical agreement with experimental results using this force field may be challenging. To address this issue, ab inito methods could be employed. 

Another factor is the refinement of fitting schemes to estimate the parameters of $\Gamma_b$. More efficient techniques could be utilized to compute the linewidths of the VDSF from larger systems without the need of explicitly diagonalizing large matrices~\cite{vast2000effects}. 

Future work will address these issues to improve the accuracy of our method and enhance the quantitative agreement with experimental results~\cite{fiorentino2023unearthing}.

\subsection*{Conclusions}

We have developed an effective model for the low-frequency excitations in amorphous solids that allows one to accurately compute the bulk limit of the thermal conductivity from glass models of moderate size. Our method, which stands on a combination of the QHGK approximation with various ideas that have been floating around in the literature for some time now, naturally accounts for the interplay of anharmonicity and disorder in determining the transport properties of glasses. The resulting protocol gets around the need for mock crystalline models for the glass, which introduce spurious modes whose group velocities bear little physical meaning. This fact is exemplified by the extreme case of a $2 \times 2 \times 2$ $q$-mesh, where group velocities vanish identically as all the points of the mesh lie on the surface of the BZ and differ by a reciprocal-lattice vector from their negatives. As a result, the thermal conductivity computed on a $2 \times 2 \times 2$ $q$-mesh is practically indistinguishable from the one computed at the zone center. Moreover, sampling the full BZ of a crystalline model for a glass only affects the first, \emph{intraband} RTA-like, term in Eq.~\eqref{eq:kappa qhgk crystal}, while the second is already at convergence at the zone center for rather small glass models. The RTA-like term grows with the size of the $q$-mesh until it converges to some value, whose magnitude may be very large and devoid of any significance.

The only requirement of our extrapolation technique is a finite sample whose size can range from hundreds up to a few thousand atoms, depending on the characteristics of the material. We have tested our model on three paradigmatic glassy materials, $\asic$, $\asio$, and $\asi$, which display different convergence properties to the bulk limit. When the effect of disorder is prominent (e.g., in amorphous silica), size effects are small, and hydrodynamic extrapolation contributes little to the converged value of $\kappa$. On the other hand, when the glass is less disordered (e.g., amorphous silicon), tens of thousands of atoms are required to obtain a converged value of the bulk thermal conductivity, while the hydrodynamic extrapolation gives satisfactory results with an order of magnitude fewer atoms.

\section*{Methods}

\subsection*{Acoustic-sound-waves basis}

For a system of $N$ atoms, $\left|\mathbf{Q},b\right\rangle$ is a $3N$-dimensional vector whose projection on the $I$th atomic site in the $\alpha$ direction is:

\begin{align}
    \langle I,\alpha|\mathbf{Q},b\rangle=\frac{1}{\sqrt{N}} \varepsilon_\alpha^b (\mathbf{Q}) e^{i \mathbf{Q}\cdot \mathbf{R}_I},
\end{align}
 
where $\bm{\varepsilon}^b (\mathbf{Q})$, with $b=L,T_1,T_2$, are three orthonormal polarization unit vectors. The scalar product between two plane-wave states is:
 
\begin{equation}
    \langle\mathbf{Q},b|\mathbf{K},b'\rangle=\frac{1}{{N}}\sum_{\alpha} \varepsilon_\alpha^{b*} (\mathbf{Q})\varepsilon_\alpha^{b'} (\mathbf{K})\sum_I e^{i (\mathbf{K}-\mathbf{Q})\cdot \mathbf{R}_I},  
\end{equation}
 
where the last sum is proportional to the Fourier Transform of the atomic number density, $\rho(\mathbf{r})=\sum_I \delta(\mathbf{r}-\mathbf{R}_I)$, i.e.
 
\begin{align}
    \begin{split}
    \tilde{\rho}(\mathbf{k}) =& \int \rho(\mathbf{r}) e^{-i \mathbf{k} \cdot \mathbf{r}} \mathrm{d}^3 r \\
    =& \sum_I \int \delta(\mathbf{r} - \mathbf{R}_I) e^{-i \mathbf{k} \cdot \mathbf{r}} \mathrm{d}^3 r \\
    =& \sum_I e^{-i \mathbf{k} \cdot \mathbf{R}_I}.
    \end{split}
\end{align}

Assuming that the material is homogeneous at length scales larger than a certain wavelength, $\lambda$, implies that ${\tilde{\rho}(\mathbf{k}<2\pi/\lambda)}$ tends to a Dirac-delta function; consequently:
 
\begin{align}
    \begin{split}
        \langle\mathbf{Q},b|\mathbf{K},b'\rangle &= \sum_{\alpha} \varepsilon_\alpha^{b*} (\mathbf{Q})\varepsilon_\alpha^{b'} \delta_{\mathbf{QK}} \\
        &= \delta_{bb'} \delta_{\mathbf{QK}}.
    \end{split}
\end{align}
 
Therefore, the subset of vectors $|\mathbf{Q},b\rangle$  with $|Q|<2\pi/\lambda$ is effectively orthonormal. Regarding the completeness problem, we are only interested in describing the propagons, not all the normal modes, which on the other hand would require a basis of $3N$ vectors. It can be argued that this set is sufficient for this purpose~\cite{taraskin2000propagation}, as it can also be qualitatively understood from the plot of the DSF, where low-frequency modes are decomposed in small-$Q$ plane waves only.

We are now in a position to write the propagon contribution to thermal conductivity in the plane-wave basis. The first term in the right-hand side of Eq.~\ref{eq:sep_kappa_prop_and_diff} in the main text couples only pairs of propagons. This is not a property of the energy flux operator, which has non-zero components for all different pairs of normal modes, but it is a consequence of the narrow Lorentzian functions appearing in the QHGK theory, Eq.~\eqref{eq:kappa qhgk}. Taking into account this separation in frequency thanks to the Heaviside step function, let us consider the energy flux operator
 
\begin{align}
    \begin{split}
        \widehat J &= \hbar\sum_{\nu \nu'}\frac{\omega_{\nu}+\omega_{ \nu'}}{2}v_{\nu\nu'}\had_{\nu} \ha_{\nu'} \Theta(\omega_\mathrm{P}-\omega_\nu) \Theta(\omega_\mathrm{P}-\omega_{\nu'}) \\
        &= \hbar\sum_{\nu \nu' \in P}
        \frac{\omega_{\nu}+\omega_{ \nu'}}{2}v_{\nu\nu'}\had_{\nu} \ha_{\nu'}
    \end{split}
\end{align}
 
where the $P$ subscript means that only pairs of propagons are involved; the anti-resonant part of the flux, which gives a negligible contribution to thermal conductivity~\cite{isaeva2019modeling}, is omitted. The generalized velocity matrix is defined as~\cite{isaeva2019modeling,barbalinardo2020efficient}:
 
\begin{align}\label{eq: generalized velocity}
    v_{\nu\nu'}^\alpha=\frac{1}{2\sqrt{\omega_\nu\omega_{\nu'}}}\sum_{IJ\beta\gamma}(R_I^\alpha-R_J^\alpha)\Phi_{I\beta}^{J\gamma}\epsilon_{I\beta}^\nu\epsilon_{J\gamma}^{*\nu'}
\end{align}
 
where $\alpha,\beta,\gamma$ are Cartesian coordinates and $\Phi_{I\beta}^{J\alpha}=\frac{1}{\sqrt{M_IM_J}}\frac{\partial^2 V}{\partial R^\alpha_I \partial R^\beta_J}$ is the Dynamical Matrix element between atoms $I$ and $J$, whose masses and positions are respectively $M_I$, $M_J$ and $\mathbf{R}_I$, $\mathbf{R}_J$, and $V$ is the potential energy. To unclutter the notation, from now on we omit the sums over Cartesian-coordinate labels, which is understood when indices are repeated. 
In order to get to $J_{\mathbf{Q}}^{bb'}$, we first expand the normal modes in plane waves
 
\begin{align*}
    \epsilon_{I\alpha}^\nu=\frac{1}{\sqrt{N}}\sum_{\mathbf{Q}b}\langle\mathbf{Q},b|\nu\rangle e^{i\mathbf{Q}\cdot R_I} \bm{\varepsilon}_\alpha^b (\mathbf{Q}),
\end{align*}
 
and we observe that for propagons:
 
\begin{align}
    \begin{split}
        \sum_{\nu}^P \omega_\nu\langle \nu|\mathbf{Q},b\rangle \ha_\nu \approx & c_bQ\sum_{\nu}^P \langle \nu|\mathbf{Q},b\rangle \ha_\nu\\
        \approx& c_bQ\ha_{\mathbf{Q}b}.
    \end{split}
\end{align}
 
The important point here is that we are able to factor $\omega_\nu$ out of the sum due to the almost-linear dispersion of the DSF. This translates into an orthogonality relation of normal modes whose frequency is far from the dispersion line, i.e., $|\langle \mathbf{Q},b|\nu\rangle|\approx 0$ if $|\omega_n-c_bQ|\ll \Gamma_b(Q)$. Plugging the last equations into the energy flux operator of the propagons yields:
 
\begin{align}
    \begin{multlined}
        \widehat{\mathbf{J}} = \frac{1}{N}\sum_{\mathbf{Q}\mathbf{K} b b'}\frac{c_bQ+c_{b'}K}{4\sqrt{c_bc_{b'}QK}}\had_{\mathbf{Q}b}\ha_{\mathbf{K}b'} \bm{\varepsilon}_\alpha^b(\mathbf{Q}) \bm{\varepsilon}_{\beta}^{b'}(\mathbf{K}) \\
        \times \sum_{IJ} (\mathbf{R}_I-\mathbf{R}_J) \Phi_{I\alpha}^{J\beta} e^{i(\mathbf{Q}\cdot\mathbf{R}_I-\mathbf{K}\cdot\mathbf{R}_J)}.    
    \end{multlined}
\end{align}
 
Under the assumption that the material is practically homogeneous above the $\lambda$ scale, the dynamical matrix can only depend on $\mathbf{R}_I-\mathbf{R}_J$; therefore, with a change of variables $(\mathbf{R}_I,\mathbf{R}_J) \mapsto (\mathbf{R}_I+\mathbf{R}_J,\mathbf{R}_I-\mathbf{R}_J)$, we obtain
 
\begin{align*}
    \frac{1}{N}\sum_{\mathbf{R}_I+\mathbf{R}_J} e^{i(\mathbf{Q}-\mathbf{K})\cdot (\mathbf{R}_I+\mathbf{R}_J)/2}=\delta_{\mathbf{Q}\mathbf{K}}  
\end{align*}
 
and then:
 
\begin{align}
    \widehat{\mathbf{J}} = -i \hbar \sum_{\mathbf{Q}b b'}\frac{c_b+c_{b'}}{4\sqrt{c_bc_{b'}}}\had_{\mathbf{Q}b}\ha_{\mathbf{Q}b'} \bm{\varepsilon}^b_\alpha(\mathbf{Q}) \bm{\varepsilon}_\beta^{b'}(\mathbf{Q}) \nabla_\mathbf{Q} \Phi_\alpha^\beta(\mathbf{Q}),
\end{align}
 
where
 
\begin{align}
    \begin{split}
        \nabla_\mathbf{Q} \Phi_\alpha^\beta(\mathbf{Q})=& \nabla_\mathbf{Q}\sum_{\mathbf{R}_I-\mathbf{R}_J}e^{i\mathbf{Q}\cdot(\mathbf{R}_I-\mathbf{R}_J)}\Phi_\alpha^\beta(\mathbf{R}_I-\mathbf{R}_J) \\
        =& \sum_{\mathbf{R}_I-\mathbf{R}_J} i e^{i\mathbf{Q}\cdot(\mathbf{R}_I-\mathbf{R}_J)} (\mathbf{R}_I-\mathbf{R}_J)\Phi_\alpha^\beta(\mathbf{R}_I-\mathbf{R}_J).
    \end{split}
\end{align}
 
From the DSF, we know that $|\mathbf{Q},b \rangle$ is practically a linear combination of almost-degenerate normal modes, that are eigenvectors of the dynamical matrix with eigenvalue $c_b^2Q^2$; therefore, for $b=b'$, we have that:
 
\begin{align}
    \sum_{\alpha,\beta}\bm\varepsilon^\alpha_b(\mathbf{Q})\bm\varepsilon^\beta_{b}(\mathbf{Q}) \nabla_\mathbf{Q} \Phi_\alpha^\beta(\mathbf{Q})\approx 2c_b^2\mathbf{Q}.
\end{align}
 
Thus, the energy flux becomes
 
\begin{align}
    \widehat{\mathbf{J}} = -i \hbar \sum_{\mathbf{Q},b} c_b^2 \had_{\mathbf{Q}b}\ha_{\mathbf{Q}b} \mathbf{Q} + \text{mixed-polarization terms}.    
\end{align}
 
The mixed-polarization terms do not contribute to the value of thermal conductivity, since the two polarizations are well separated in frequency. As such, these terms are not included in our calculations.

\subsection*{Melt-and-quench generation of glass samples}

Amorphous samples are obtained through a melt-and-quench procedure starting from a crystalline conventional cubic cell replicated $\ell$ times along each Cartesian direction. The molecular dynamics simulations are carried out using the Large-scale Atomic/Molecular Massively Parallel Simulator (LAMMPS~\cite{LAMMPS}). The melt-and-quench procedures are performed according to the following recipes:
 
\begin{itemize}
    \item[$\asio$.] The interatomic forces are described with the Vashishta force-field~\cite{vashishta1990interaction}. Simulations are carried out with a timestep of $1\,\mathrm{fs}$. Starting from the $\beta$-cristobalite cubic conventional unit cell with a mass density of $2.20,\mathrm{g~cm^{-3}}$, the crystal is melted at $7000\,\mathrm{K}$. The molten sample is quenched from $7000\,\mathrm{K}$ to $500\,\mathrm{K}$ in $10\,\mathrm{ns}$~\cite{ercole2017accurate, ercole2018ab}. The sample is then thermalized at $500\,\mathrm{K}$ for $400\,\mathrm{ps}$, and finally equilibrated for $100\,\mathrm{ps}$ in the $NVE$ ensemble. For each size, we randomized the seed of the thermostat to obtain ten different samples. The average mass density of the amorphous samples is $2.43\,\mathrm{g~cm^{-3}}$, with a standard deviation across sizes of $0.01\,\mathrm{g~cm^{-3}}$.
    \item[$\asic$.] The interatomic forces are described with the Vashishta force-field~\cite{vashishta2007interaction,vashishta2004short}. Simulations are carried out with a timestep of $1\,\mathrm{fs}$. For each size, the starting configuration is a crystalline cubic zinc-blend structure with a mass density of $3.22\,\mathrm{g~cm^{-3}}$.  Following Ref.~\onlinecite{vashishta2007interaction}, the system is gradually heated from $300\,\mathrm{K}$ to $4000\,\mathrm{K}$ at constant null pressure. The solid/liquid transition is characterized by a sharp increase in the volume between $3000$ and $3500\,\mathrm{K}$~\cite{vashishta2007interaction}. Ten different liquid configurations are extracted every $2\,\mathrm{ps}$ from an $NVE$ trajectory. The molten configurations are then quenched to $1000\,\mathrm{K}$ in $300\,\mathrm{ps}$, and thermalized for $80\,\mathrm{ps}$ at constant pressure. The system is further cooled down to $500\,\mathrm{K}$ with the same procedure. The average mass density of the amorphous samples is $2.98\,\mathrm{g~cm^{-3}}$, with a standard deviation across sizes of $0.01\,\mathrm{g~cm^{-3}}$.
    \item[$\asi$.] The interatomic forces are described with the Tersoff force-field~\cite{tersoff1988empirical}. Simulations are carried out with a timestep of $0.5\,\mathrm{fs}$. For each size, the starting configuration is a crystalline diamond conventional unit cell replicated $\ell$ times along each Cartesian direction. Following Ref.~\onlinecite{deringer2018realistic}, the crystal is melted at $6000\,\mathrm{K}$ and brought to $3000\,\mathrm{K}$ in $2\,\mathrm{ns}$ at fixed zero pressure. Then the system is equilibrated at $3000\,\mathrm{K}$ for another $2\,\mathrm{ns}$. The molten sample is successively quenched from $3000\,\mathrm{K}$ to $2000\,\mathrm{K}$ in $10\,\mathrm{ns}$, and finally annealed from $2000\,\mathrm{K}$ to $300\,\mathrm{K}$ at fixed volume in another $10\,\mathrm{ns}$. Each glassy sample is equilibrated at $300\,\mathrm{K}$ for $10\,\mathrm{ns}$. For each size, ten different samples are prepared according to this recipe beginning the quenching procedure from liquid configurations obtained initializing the atomic velocities with different random seeds. The average mass density of the amorphous samples is $2.275\,\mathrm{g~cm^{-3}}$, with a standard deviation across sizes of $0.003\,\mathrm{g~cm^{-3}}$.
\end{itemize}

\subsection*{Computation of the dynamical structure factors}

The dynamical structure factors (DSFs) are computed according to Eq.~\eqref{eq:dsf anharmonic} in the main text. For low frequencies, the anharmonic linewidths become smaller than the finite-size frequency spacing. When this happens, the Lorentzian functions in Eq.~\eqref{eq:dsf anharmonic} are so narrow that few to no data points fall within its width. Thus, the contribution to the DSF coming from low-frequency modes is undersampled and affected by numerical noise. Under the assumption that the DSF is a Lorentzian function, one can overcome this issue by exploiting the closure of Lorentzian functions under convolutions. In fact, introducing the shorthand notation $A_\nu(\mathbf{Q}) \equiv |\langle \nu| \mathbf{Q},b\rangle|^2$, and

\begin{align*}
    \mathcal{L}(\omega, \eta) \equiv \frac{1}{\pi} \frac{\eta}{\omega^2 + \eta^2},
\end{align*}
 
a modified version of the DSF can be introduced as:
 
\begin{align}\label{eq:trick dsf}
    \begin{split}
        S^\eta_b(\omega,\mathbf{Q}) &= \sum_{\nu} |\langle \nu| \mathbf{Q},b\rangle|^2 \frac{1}{\pi}\frac{\gamma_\nu+\eta}{(\gamma_\nu + \eta)^2+(\omega-\omega_\nu)^2} \\
        &= \sum_\nu A_n(\mathbf{Q}) \mathcal{L}(\omega-\omega_\nu, \gamma_\nu+\eta)
    \end{split}
\end{align}
 
Then, by the closure of Lorentzians under convolution:
 
\begin{align}
    \begin{split}
        S^\eta_b(\omega,\mathbf{Q})&= \int \mathrm{d}\omega'  \sum_{\nu} A_n(\mathbf{Q}) \mathcal{L}(\omega', \gamma_\nu) \mathcal{L}(\omega'-(\omega-\omega_\nu), \eta) \\
        &= \int \mathrm{d}\omega'  \sum_{\nu} A_n(\mathbf{Q}) \mathcal{L}(\omega'+\omega-\omega_\nu, \gamma_\nu) \mathcal{L}(\omega', \eta) \\
        &= \int \mathrm{d}\omega' \mathcal{L}(\omega', \eta) S_b(\omega,\mathbf{Q}).
    \end{split}
\end{align}
 
Since, for propagons, we assume that $S_b(\omega,\mathbf{Q})$ is a Lorentzian function centered on $\omega_b(\mathbf{Q}) = c_b Q$ and whose width is $\Gamma_b(Q)$, we get
 
\begin{align}
    \begin{split}
        S^\eta_b(\omega,\mathbf{Q}) &= \int \mathrm{d}\omega' \mathcal{L}(\omega', \eta) \mathcal{L}(\omega'+\omega-c_b Q, \Gamma_b(Q)) \\
        &= \mathcal{L}(\omega- c_b Q, \Gamma_b(Q)+\eta).
    \end{split}
\end{align}
 
Due to this property, one can compute $\Gamma_b(Q)$ subtracting $\eta$ from the width of $S^\eta_b(\omega,\mathbf{Q})$ numerically calculated from Eq.~\eqref{eq:trick dsf} with a large-enough $\eta$, since $S^\eta_b(\omega,\mathbf{Q})$ is much less affected by undersampling than $S_b(\omega,\mathbf{Q})$.

\subsection*{Interpolation scheme for the anharmonic linewidths}

The computation of third-order anharmonic linewidths in glasses generally constitutes the bottleneck of QHGK calculations~\cite{barbalinardo2020efficient}, as it scales as $N_{\mathrm{atoms}}^3$. Thus, this computation by itself would severely limit our ability to study size effects. It is thus customary to adopt some interpolation scheme for obtaining the linewidths of a large model starting from a smaller one~\cite{isaeva2019modeling, barbalinardo2020efficient, simoncelli2023thermal}.
First, in order to smoothen the data, we apply a Gaussian filter to the computed anharmonic linewidths at each temperature, $\gamma_\nu(T)$:
 
\begin{align}
    \gamma(\omega,T)=\frac{\sum_\nu \gamma_\nu (T)\frac{1}{\sqrt{2\pi\sigma^2}} \exp[\frac{(\omega-\omega_\nu)^2}{2\sigma^2}]}{\sum_\nu  \frac{1}{\sqrt{2\pi\sigma^2}} \exp[\frac{(\omega-\omega_\nu)^2}{2\sigma^2}]}.  
\end{align}
 
Then, we spline-interpolate the smoothed function imposing a quadratic behavior at vanishing frequencies, which is compatible with the hydrodynamics of solids~\cite{griffin1968brillouin}. We average $\gamma(\omega,T)$ over disorder by computing the mean of the interpolated linewidths over different same-size samples of each material. The spline functions are finally evaluated on the frequencies of larger samples in order to obtain their approximated anharmonic linewidths.

\section*{Data Availability}
The data that support the plots and relevant results within this paper are available on the Materials Cloud platform~\cite{talirz2020materials}. See DOI:\href{https://archive.materialscloud.org/record/2023.120}{10.24435/materialscloud:k2-0n}.

\section*{Code Availability}
The codes that support the relevant results within this paper are available to the respective developers. Analysis scripts are available on GitHub~\cite{Fiorentino_hydro_glasses_2023} and on the Materials Cloud platform~\cite{talirz2020materials}. See DOI:\href{https://archive.materialscloud.org/record/2023.120}{10.24435/materialscloud:k2-0n}. 

\section*{Acknowledgements}

We thank E. Drigo, M. G. Izzo, G. Tenti, and G. Barbalinardo for fruitful discussions. We also thank F. Grasselli for a critical reading of the manuscript. This work was partially funded by the EU through the \textsc{MaX} \emph{Centre of Excellence for supercomputing applications} (Project No. 10109337), by the Italian Ministry of Research and Education through by the Italian MUR through the PRIN 2017 \emph{FERMAT} (grant No. 2017KFY7XF), and by the \emph{National Centre from HPC, Big Data, and Quantum Computing} (grant No. CN00000013). 

\section*{Competing interests}

The authors declare no competing interests.

\section*{Author contribution}

AF and PP are to be considered co-first authors. SB supervised the work.

\newpage
\bibliography{main}

\end{document}